\documentclass[twocolumn, twocolappendix]{aastex631}
\usepackage{amsmath}
\usepackage{mathrsfs}
\usepackage{graphicx}
\usepackage{xcolor}
\usepackage{dcolumn}
\usepackage{bm}
\usepackage{amssymb}
\usepackage{amsmath}
\usepackage{verbatim}
\usepackage[utf8]{inputenc}
\usepackage{tikz,hyperref}
\usepackage{dcolumn}

\shortauthors{Yang, Zhang, Bi, \& Yin}

\begin{document}

\title{Interpreting the Extremely Diffuse Stellar Distribution of the Nube Galaxy through Fuzzy Dark Matter}

\author[0009-0005-5375-9437]{Yu-Ming Yang}
\email{yangyuming@ihep.ac.cn}
\affiliation{Key Laboratory of Particle Astrophysics, Institute of High Energy Physics, Chinese Academy of Sciences, Beijing 100049, China}
\affiliation{University of Chinese Academy of Sciences, Beijing 100049, China }

\author[0009-0004-1366-1294]{Zhao-Chen Zhang}
\email{zhangzhaochen@ihep.ac.cn}
\affiliation{Key Laboratory of Particle Astrophysics, Institute of High Energy Physics, Chinese Academy of Sciences, Beijing 100049, China}
\affiliation{University of Chinese Academy of Sciences, Beijing 100049, China }

\author[0000-0002-5334-9754]{Xiao-Jun Bi}
\email{bixj@ihep.ac.cn}
\affiliation{Key Laboratory of Particle Astrophysics, Institute of High Energy Physics, Chinese Academy of Sciences, Beijing 100049, China}
\affiliation{University of Chinese Academy of Sciences, Beijing 100049, China }

\author[0000-0001-6514-5196]{Peng-Fei Yin}
\email{yinpf@ihep.ac.cn}
\affiliation{Key Laboratory of Particle Astrophysics, Institute of High Energy Physics, Chinese Academy of Sciences, Beijing 100049, China}

\begin{abstract}
Recent observations have uncovered a remarkably flat and extremely diffuse stellar distribution within the almost dark dwarf galaxy Nube, posing a challenge to the standard cold dark matter  scenario. In this study, we employ numerical simulations to explore the possibility that this anomalous stellar distribution can be attributed to the dynamical heating effect of fuzzy dark matter (FDM). The relatively isolated location and low baryon fraction of Nube make it an ideal system for investigating this effect. Our findings indicate that by adopting a halo profile consistent with the dynamical mass estimation of Nube and an FDM particle mass on the order of $10^{-23}$ eV, the final 2D stellar distribution derived from simulation closely matches observational data. These results suggest that FDM could provide an explanation for the extremely diffuse stellar distribution of Nube.

\end{abstract}

\keywords{
\href{http://astrothesaurus.org/uat/353}{Dark matter (353)};
\href{http://astrothesaurus.org/uat/940}{Low surface brightness galaxies(940)};
\href{http://astrothesaurus.org/uat/1880}{Galaxy dark matter halos (1880)}
}

\section{Introduction}

An almost dark galaxy was fortuitously discovered in the IAC Stripe 82 Legacy Project \citep{Fliri_2015}. Recent observational analyses using data from the 100 m Green Bank Telescope and the 10.4 m Gran Telescopio Canarias revealed that this galaxy, named Nube, has a total stellar mass of around $3.9\,\times\,10^8\,M_\odot$, a HI to stellar mass ratio of around $1$, and a much larger dynamical mass within $~20.7$ kpc, estimated to be about $2.6\,\times\,10^{10}\,M_\odot$ \citep{montes2024almost}. However, its surface stellar density distribution deviates significantly from that of other dwarf galaxies, exhibiting a notably flatter profile and an extremely low central density ($\sim 2\,M_\odot\,\text{pc}^{-2}$). Moreover, the effective radius of Nube surpasses even that of ultradiffuse galaxies (UDGs) with comparable stellar masses \citep{Chamba_2020}. 

These features indicate that the density of dark matter (DM) is at least approximately an order of magnitude higher than the baryonic matter at all locations in Nube, hence the impact of baryonic effects \citep{Governato:2009bg} such as feedback is minimal. Furthermore, Nube is situated in a relatively isolated position, at a projected distance of approximately 435 kpc from its most likely host halo, UGC 929. Observations of the morphology and surrounding environment of Nube suggest that this galaxy has not experienced strong tidal distortions \citep{montes2024almost}. These distinctive characteristics pose challenges for explaining the origin of Nube within the framework of cold dark matter (CDM). In the CDM framework, isolated galaxies with low baryon fraction tend to have stellar distributions that are more centrally concentrated. Galaxies with properties similar to Nube have not been identified in CDM simulations that successfully reproduce the characteristics of the largest known UDGs \citep{montes2024almost}. Therefore, the characteristics of Nube imply that the nature of DM may deviate from CDM.

In this study, we demonstrate that the diffuse stellar distribution of Nube can be explained in the scenario of fuzzy dark matter \citep{hu2000fuzzy,peebles2000fluid,hui2017ultralight,hui2021wave} through numerical simulations. The dynamical heating effect \citep{Bar_Or_2019,Dutta_Chowdhury_2021,Dutta_Chowdhury_2023} in a FDM halo can transfer energy to the stars, resulting in a diffuse stellar distribution \citep{Yang:2024vgw}. We utilize the eigenstate decomposition method \citep{Yavetz_2022,Alvarez-Rios:2024vhs} to construct the initial wave function of FDM within the halo, and employ the \textsc{PyUltraLight} package \citep{Edwards_2018}, which adopts the pseudospectral method, to evolve the wave function satisfying the Schr$\ddot{\text{o}}$dinger-Poisson (SP) equations. In our simulations, stars are treated as massless particles,  given that the gravitational field in Nube is predominantly governed by DM. The stars are initialized based on the Plummer profile \citep{1911MNRAS..71..460P} and the Eddington formula \citep{10.1093/mnras/76.7.572}, and evolve within the gravitational potential of FDM. A diffuse stellar distribution, consistent with observational data, emerges after 10.2 Gyr, corresponding to the estimated age of Nube, for a $m_a$ on the order of  $10^{-23}$ eV.

This Letter is organized as follows. In Section \ref{Sec2}, we outline our simulation setup. We then present the simulation results of Nube and compare them to observational data in Section \ref{Sec3}. In Section \ref{Sec4}, we discuss the implications of our results and conclude our study. Additional details on the construction of the FDM halo and the evolution of our systems are provided in Appendices \ref{App_A} and \ref{App_B}, respectively.

\section{Simulation Setup\label{Sec2}}
\subsection{FDM Halo Construction}

\begin{figure}[htbp]
    \centering
    \includegraphics[width=\linewidth]{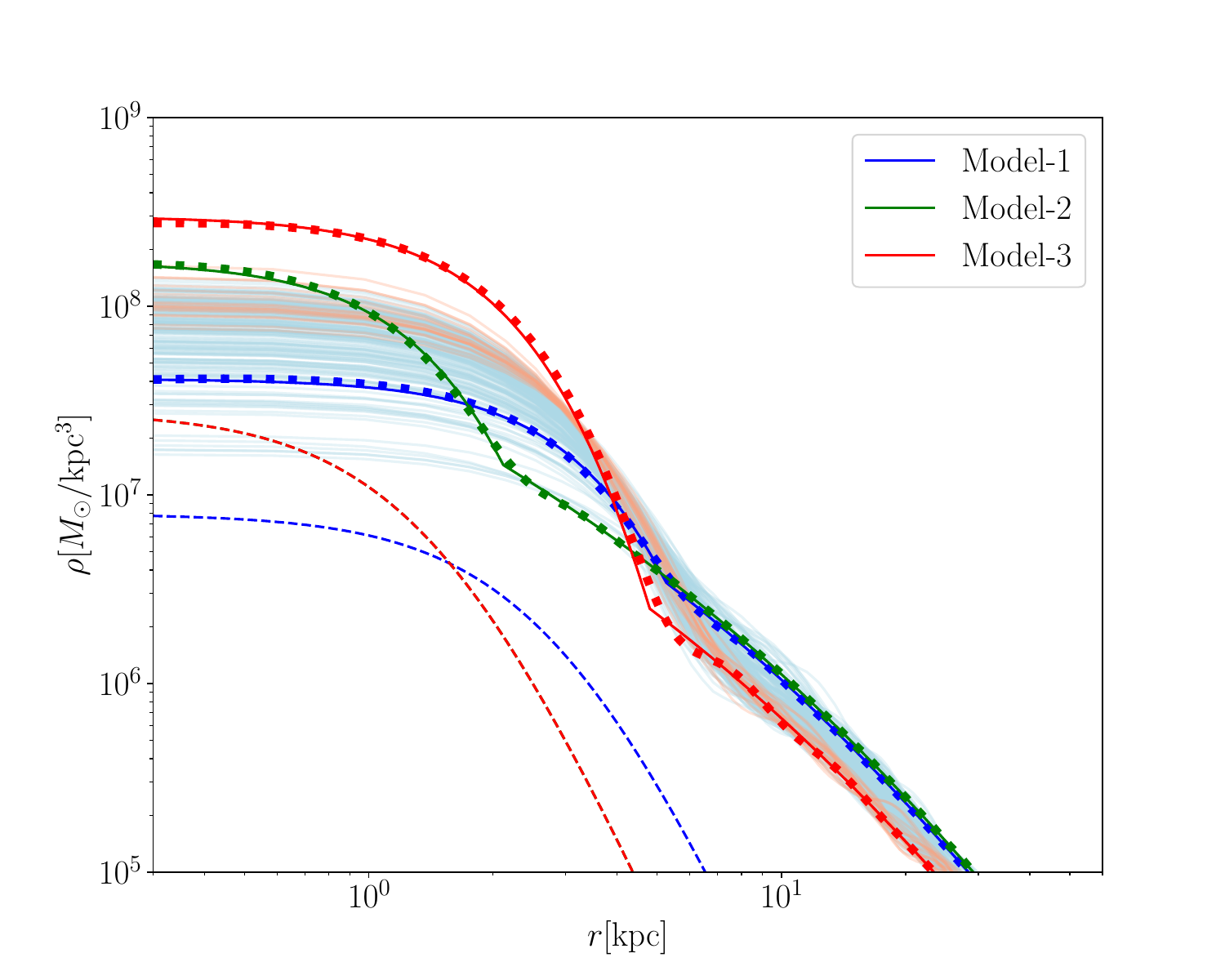}
    \caption{
    Radial FDM profiles for the three models under consideration are shown. The blue, green, and red solid lines represent the target FDM profiles $\rho_\text{in}(r)$ used as input for generating the initial wave functions. The squares represent the reproduced density profiles $\rho_\text{out}(r)$ obtained from the derived initial wave functions. The dashed lines represent the initial stellar density profiles. Note that the lines representing the stellar density profiles of Model-2 and Model-3 overlap in the figure. The light blue and light orange lines represent the spherical-averaged FDM density profiles of Model-1 at various snapshots during the first 9.2 Gyr and final 1 Gyr of evolution, respectively. Each snapshot is separated by a 50 Myr interval.}
    \label{model}
\end{figure}

In the nonrelativistic limit, FDM can be described as a classical field $\psi(t,\mathbf{x})$, which obeys the SP equations \citep{hui2021wave}
\begin{equation}
    \begin{aligned}
        i\hbar \partial_t \psi&=-\frac{\hbar^2}{2m_a}\bm{\nabla}^2\psi+m_a\Phi\psi,\\
        \bm{\nabla}^2\Phi&=4\pi G\rho, \quad \rho=m_a|\psi|^2,
    \end{aligned}
\end{equation}
where $m_a$ is the mass of the particle, $\Phi$ is the gravitational potential, and $\rho$ is the mass density. Since the stellar mass of Nube constitutes less than a few percent of the total mass, its contribution to the gravitational field has been neglected in the SP equations. Studies based on cosmological simulations \citep{Schive_2014_1,Schive_2014} have indicated that FDM halos exhibit a solitonic core representing the ground state solution of the SP equations and an Navarro-Frenk-White-(NFW) like envelope composed of excited states. Hence, we refer to $\psi$ as the wave function of FDM, and construct the initial $\psi$ at $t=0$ for our simulation based on a target profile consisting of a solitonic core and an NFW-like envelope 
\begin{equation}
    \rho_\text{in}(r)=\left\{\begin{aligned}
        &\frac{\rho_c}{\left[1+0.091(r/r_c)^2\right]^8},\quad r<kr_c\\
        &\frac{\rho_s}{(r/r_s)\left(1+r/r_s\right)^2},\quad r\geq kr_c,
        \end{aligned}\right.
        \label{ini_profile}
\end{equation}
where $k$ is a parameter describing the transition radius with varying values in different studies \citep{Mocz_2017,Dutta_Chowdhury_2021,Chiang_2021}. Due to the scaling symmetry of the SP equations \citep{Guzman_2006}, the core density $\rho_c$ and core radius $r_c$ are related by $\rho_c=1.95\times 10^7 M_\odot \text{kpc}^{-3}\left(m_a/10^{-22}\text{eV}\right)^{-2}\left(r_c/\text{kpc}\right)^{-4}$. In this study, we fix $r_s$ at 10 kpc, as its impact is determined to be negligible \citep{Yang:2024vgw}. Therefore, according to the continuity condition at $kr_c$, the dynamical mass of Nube within $20.7$ kpc \citep{montes2024almost}, and the scaling relation between $\rho_c$ and $r_c$, the halo density profile is determined by a set of parameters $m_a$ and $k$. In this study, we consider three sets of $m_a$ and $k$, as outlined in Table \ref{Tab1}, with the corresponding profiles depicted in Figure \ref{model} using blue, green, and red solid lines, respectively.
\begin{table}[htbp]
    \setlength{\tabcolsep}{15pt} 
    \centering
    \caption{Parameters Considered for Our Simulations}
    \begin{tabular}{ccccc}
    \hline
    \hline
    Model&$m_a$&$k$&$M_\star$&$r_\star$\\
    \hline
    Model-1& 1&2&8.9&3.0\\
    Model-2& 3&2&3.9&1.5\\
    Model-3& 1&3&3.9&1.5\\
    \hline
    \end{tabular}
    \label{Tab1}
    \tablecomments{Columns from left to right: model name label, FDM particle mass $m_a(10^{-23}\,\text{eV})$, parameter $k$ which describes the transition radius in the FDM halo, total stellar mass $M_\star(10^8M_\odot)$, and initial effective radius $r_\star(\text{kpc})$. }
\end{table}

The initial wave function utilized in our simulation is expressed as a linear combination of eigenstates \citep{Yavetz_2022}
\begin{equation}
\psi(0,\mathbf{x})=\sum_{nlm}|a_{nl}|e^{i\phi_{nlm}}\Psi_{nlm}(\mathbf{x}),
\label{psi}
\end{equation} 
where $\Psi_{nlm}(\mathbf{x})=R_{nl}(r)Y^m_l(\theta,\varphi)$ are the products of the radial wave functions and spherical harmonic functions. These $\Psi_{nlm}$ are the eigenstates of the time-independent Schr$\ddot{\text{o}}$dinger equation under the static potential $\Phi_\text{in}(r)$, which is determined by the target profile $\rho_\text{in}(r)$. The integers $n,l$, and $m$ correspond to the number of nodes in $R_{nl}$, angular, and magnetic quantum numbers, respectively. The coefficients $|a_{nl}|$ are adjusted to ensure that the random phase averaged profile $\rho_\text{out}(r)$, which is derived from $|\psi(0,\mathbf{x})|^2$, aligns with the desired input profile $\rho_\text{in}(r)$. The phases $\phi_{nlm}$ are randomly sampled from the interval $[0,2\pi)$. Further details on the techniques employed in constructing the halo, including the methodology for obtaining $\Psi_{nlm}(\mathbf{x})$ and $|a_{nl}|$, are provided in Appendix \ref{App_A}. The output profiles $\rho_\text{out}(r)$ of FDM derived from the constructed $\psi(0,\mathbf{x})$ for the three models are illustrated in Figure \ref{model} as squares. It is evident that the constructed $\psi(0,\mathbf{x})$ well reproduce the target profiles. The constructed halo may exhibit an undesired nonzero global velocity attributed to the unconstrained phases introduced in the initial wave function \citep{Yang:2024trr}. To eliminate this global velocity, a Galilean boost is implemented on the wave function \citep{Yang:2024trr}.

\subsection{Stellar Initial Condition}

To incorporate the stellar component within Nube, we adopt a Plummer profile \citep{1911MNRAS..71..460P} $\rho_\star(r)=(3M_\star/4\pi r_\star^3)(1+r^2/r_\star^2)^{-5/2}$ to describe the initial stellar density distribution, 
where $M_\star$ and $r_\star$ represent the total stellar mass and initial effective radius of Nube, respectively. We investigate two parameter sets of Nube falling within the scatter range of the effective radius-stellar mass relation \citep{Chamba_2020} for typical dwarf galaxies, as outlined in Table \ref{Tab1}. The corresponding profiles are depicted in Figure \ref{model} by the dashed lines, with Model-2 and Model-3 overlapping. Our simulations reveal that the values of $r_\star$ within the range of $1.5-3.0$ kpc have minimal impact on the final 2D stellar distribution. Setting $r_\star$ to 3.0 kpc in Model-1 guarantees that the initial stellar density is significantly lower than that of DM across all radial distances, as visually observed in  Figure \ref{model}.

We utilize $10^5$ particles to represent the stellar component. This number of particles is considered sufficient, as it can produce relatively smooth 2D stellar density profiles.
The acceptance-rejection method is utilized for the Monte Carlo sampling of these particles' initial position and velocity vectors.
The initial position vectors of stars are sampled according to the Plummer profile.
To obtain a stable equilibrium system as the initial condition, the velocity vectors of star particles are sampled according to the isotropic distribution function $f(\mathcal{E})$. $f(\mathcal{E})$ is numerically computed using the Eddington formula \citep{10.1093/mnras/76.7.572}
\begin{equation}
    f(\mathcal{E})=\frac{1}{\sqrt{8}\pi^{2}}\frac{\rm d}{{\rm d}\mathcal{E}}\int_{0}^{\mathcal{E}}\frac{{\rm d}\Phi_0}{\sqrt{\mathcal{E}-\Phi_0}}\frac{{\rm d}\rho_\star}{{\rm d}\Phi_0},
\end{equation}
where $\mathcal{E}$ is the energy per unit mass of the star particle, $\rho_\star$ is the stellar density, and $\Phi_0$ represents the initial gravitational field. It is assumed that $\Phi_0$ is equal to $\Phi_\text{out}(r)$ solely determined by $\rho_\text{out}(r)$, as the contribution of the stellar component can be neglected.

To ensure that the stellar component reaches thermal equilibrium under the initial gravitational potential $\Phi_\text{out}(r)$, we conduct a verification test. Initially, we evolve the $10^5$ particles in $\Phi_\text{out}(r)$ for approximately 2 Gyr. After this initial evolution, we take the star particles that persist within the simulation box as the actual initial condition, resulting in a slightly reduced (by less than $5\%$) number of star particles compared to $10^5$. To test the stability of the initial condition, we evolve these remaining particles in $\Phi_\text{out}(r)$ for a period of 10.2 Gyr. The simulation results show that the stellar distribution remains stable over this period, with the velocity dispersion maintaining isotropy throughout the evolution.
\begin{figure*}[htbp]
  \centering
  \includegraphics[width=0.625\textwidth]{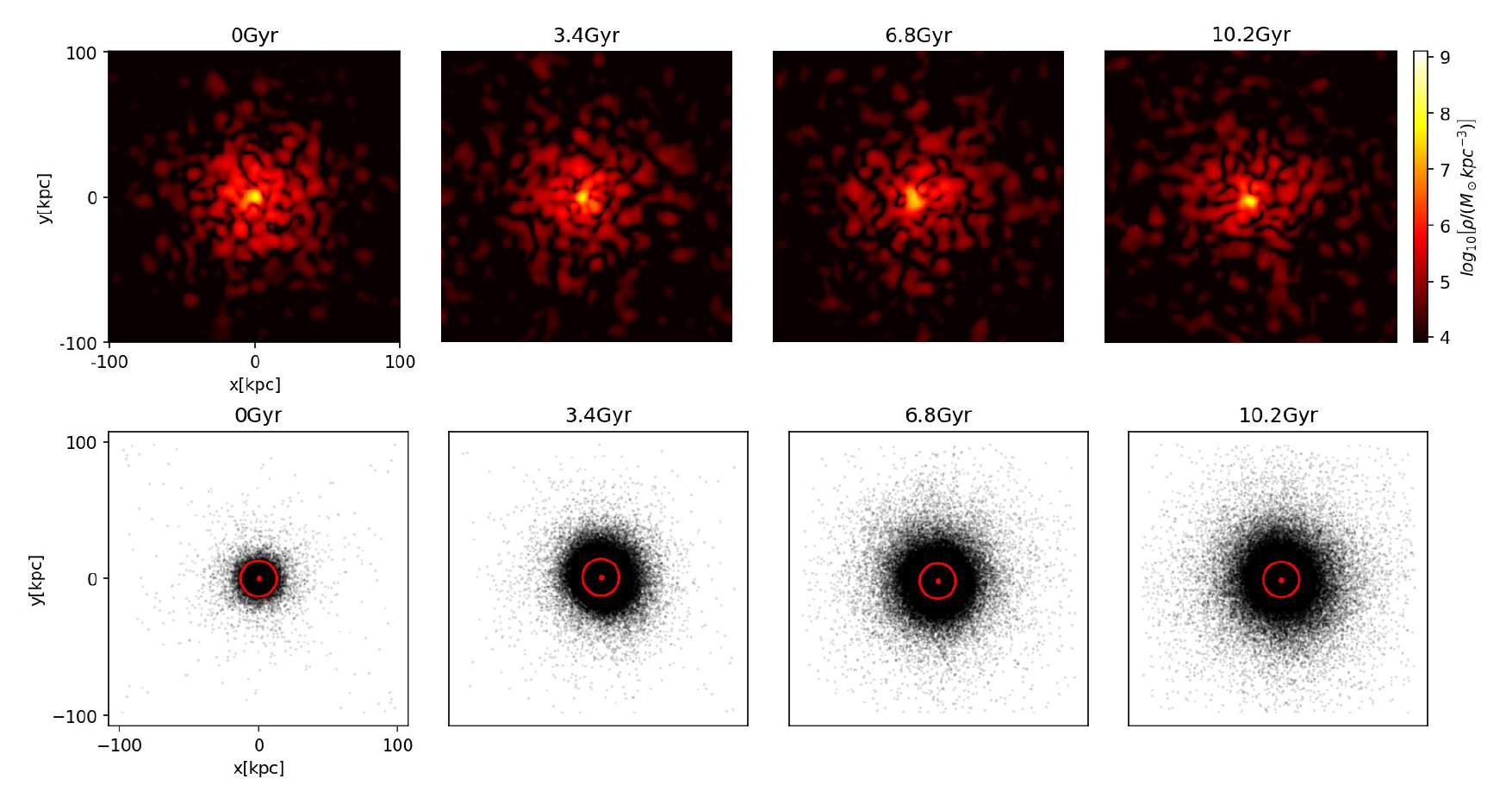}
  \includegraphics[width=0.355\textwidth]{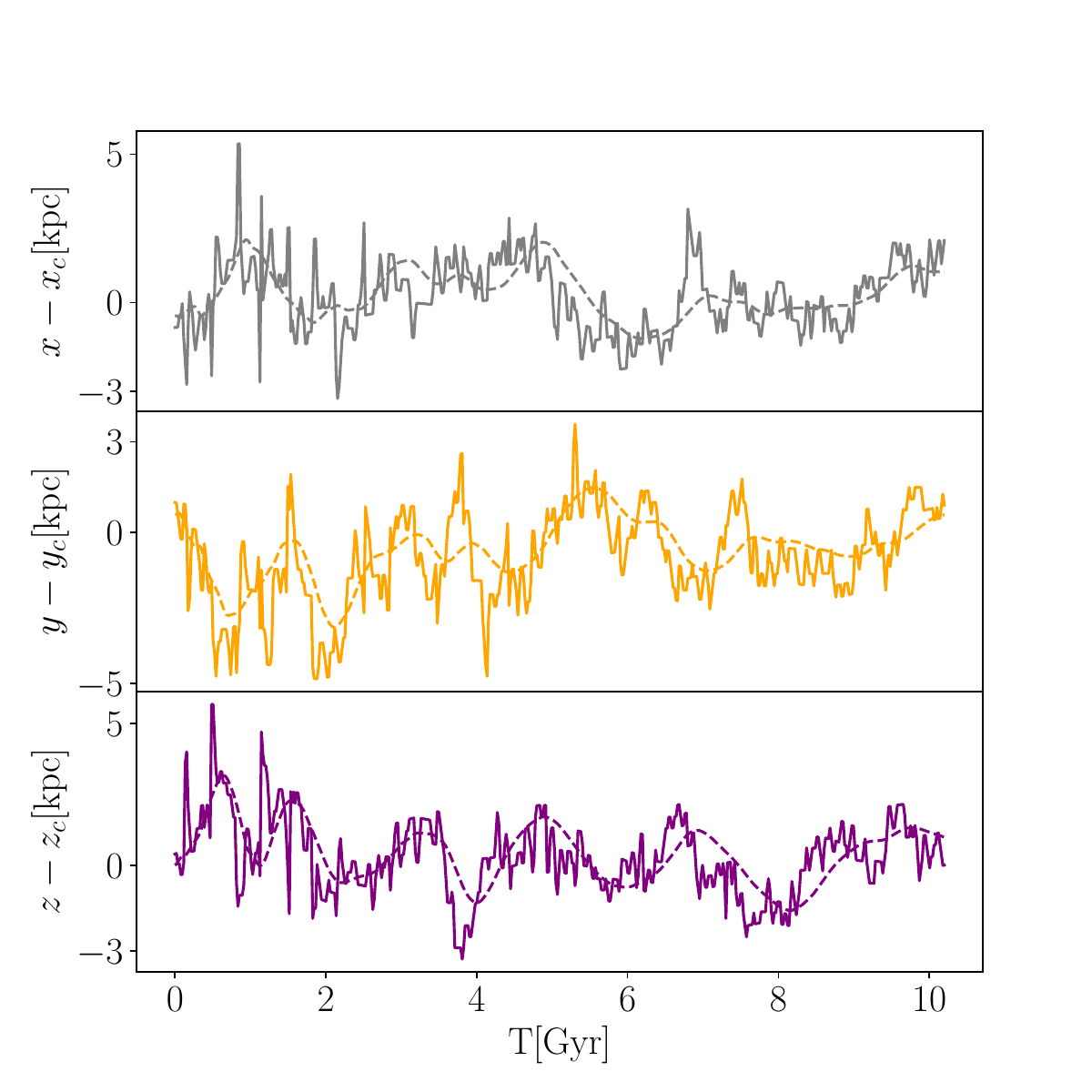}
  \caption{Left panel: FDM density field $\rho=m_a|\psi|^2$ in the $z=0$ plane (top row) and the projected positions of the star particles onto the x-y plane (bottom row) at four snapshots throughout the entire simulation duration. The red dots and circles in the bottom row represent the position of the stellar mass center and the locations at a distance of $R=13$ kpc from the mass center, corresponding to the maximum observational range. Right panel: relative coordinate of the soliton center and stellar mass center concerning the halo mass center throughout the simulation duration. Gray, orange, and purple distinguish the x, y, and z coordinates, while solid and dashed lines differentiate between the soliton and the stellar component. All the results in this figure are derived from the analysis of Model-1.}
  \label{evolution}
\end{figure*}

\subsection{Evolution of the System}

The evolution of the FDM wave function is carried out using the \textsc{PyUltraLight} package,  which employs the pseudospectral method described in \cite{Edwards_2018}, with a concise summary provided in Appendix \ref{App_B}. We apply periodic boundary conditions within a $(200~\text{kpc})^3$ simulation box with a resolution of $512^3$. Based on the cosmological parameters derived from Planck 2018 \citep{Planck_2018}, the virial radii $r_{200}$ at $z=0$ in our three models are approximately 89.2, 90.9, and 79.8 kpc, respectively. Therefore, throughout the entire evolution process, the virial radius is less than half the length of one side of the simulation box. Furthermore, it has been verified that our simulation outcomes remain consistent even with higher resolutions or larger region box lengths. During the evolution, the stars are treated as massless test particles and their gravitational influence on the system is neglected, as the gravitational field is predominantly governed by DM. The fourth-order Runge-Kutta integrator is employed to evolve the motion of star particles within the gravitational potential. Any star particles that exceed the boundaries of the simulation box during the evolution process are removed from the simulation.

The time steps for the evolution of the FDM wave function and star particles are set to be $\Delta t_\text{FDM}=0.971~\text{Myr}$ and $\Delta t_\star=0.097~\text{Myr}$, respectively. Consequently, after one step of evolution of the FDM wave function, the star particles undergo evolution for 10 steps. It is assumed that during these 10 steps of stellar evolution, the density distribution of FDM and the corresponding gravitational field remain unchanged over time. It has been verified that the results remain stable with smaller time steps. Each simulation is conducted for a duration of 10.2 Gyr, corresponding to the age of Nube. Further details on the system's evolution can be found in Appendix \ref{App_B}.

\section{Results\label{Sec3}}
\subsection{FDM and Stellar Motion}

We use Model-1 as an example to demonstrate the motion of FDM and stars in our simulations. In the top row of the left panel of Figure \ref{evolution}, we present the FDM density $\rho=m_a|\psi|^2$ in the $z=0$ plane at four different snapshots. This visualization captures the dynamic behavior of the structures within the system, showcasing the soliton as a concentrated, dense region at the core, surrounded by granules that exhibit a more diffuse and evolving fluctuating distribution. In the right panel of Figure \ref{evolution}, we illustrate the soliton random walk effect. In this depiction, the gray, orange, and purple solid lines represent the relative x, y, and z coordinates of the soliton center (defined as the location of the densest cell) concerning the center of mass of the halo throughout the simulation duration. Meanwhile, the soliton undergoes an oscillation effect, as illustrated by the light blue and light orange lines in Figure \ref{model}. All of these dynamic structures emerge from the interference between different states \citep{Li_2021,Liu_2023,Veltmaat_2018,Schive_2020}. Another characteristic of the FDM evolution is the gradual central concentration of the spherical-averaged profile over time. This can be clearly seen in Figure \ref{model}, where the light orange lines, representing the final 1 Gyr of evolution, show an overall higher central density compared to the blue solid line, which represents the input target profile of Model-1. This trend may result from the collapse from an excited state to the ground state. The increasing central concentration of the density profile would deepen the gravitational potential, thereby strengthening the binding of the stars.

The fluctuations in the FDM density field result in corresponding fluctuations in the gravitational field, which in turn affect the distribution of star particles in the system \citep{Bar_Or_2019}.  In the bottom row of the left panel of Figure \ref{evolution}, we depict the evolution of the projected positions of the star particles on the x-y plane over time. This visualization highlights the expansion of the stellar distribution driven by the dynamical heating effect. In the right panel of Figure \ref{evolution}, the gray, orange, and purple dashed lines represent the relative x, y, and z coordinates of the stellar mass center concerning the halo mass center, respectively. This illustration intuitively demonstrates that despite the presence of some relative motion between the center of mass of the stars and the soliton center \citep{Dutta_Chowdhury_2023}, their movements generally align. This behavior emerges as a natural consequence of the dominant influence of FDM in shaping the gravitational potential, as the soliton center represents the minimum point of the gravitational potential.

\begin{figure*}[htpb]
    \centering
    \includegraphics[width=0.48\linewidth]{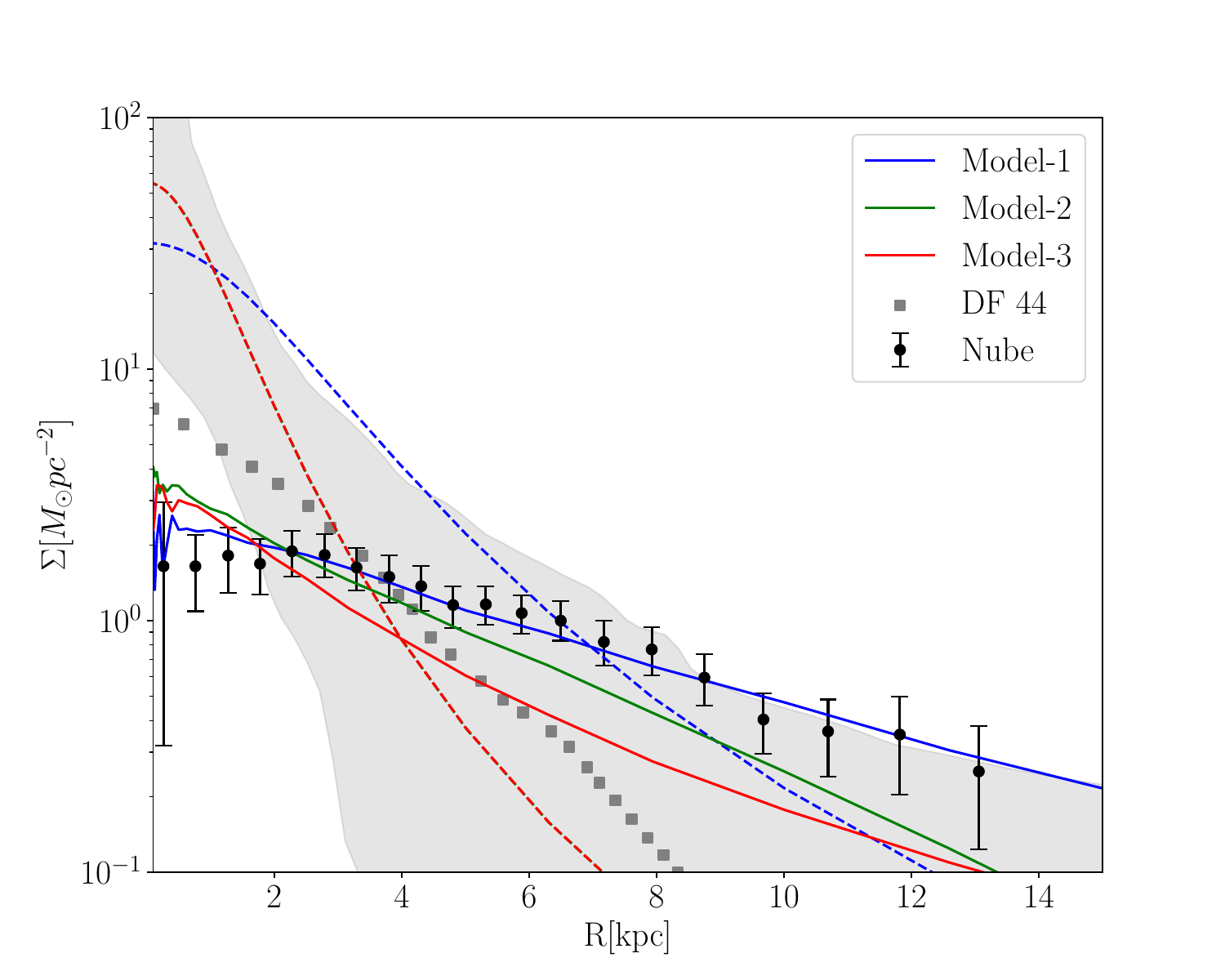}
    \includegraphics[width=0.48\linewidth]{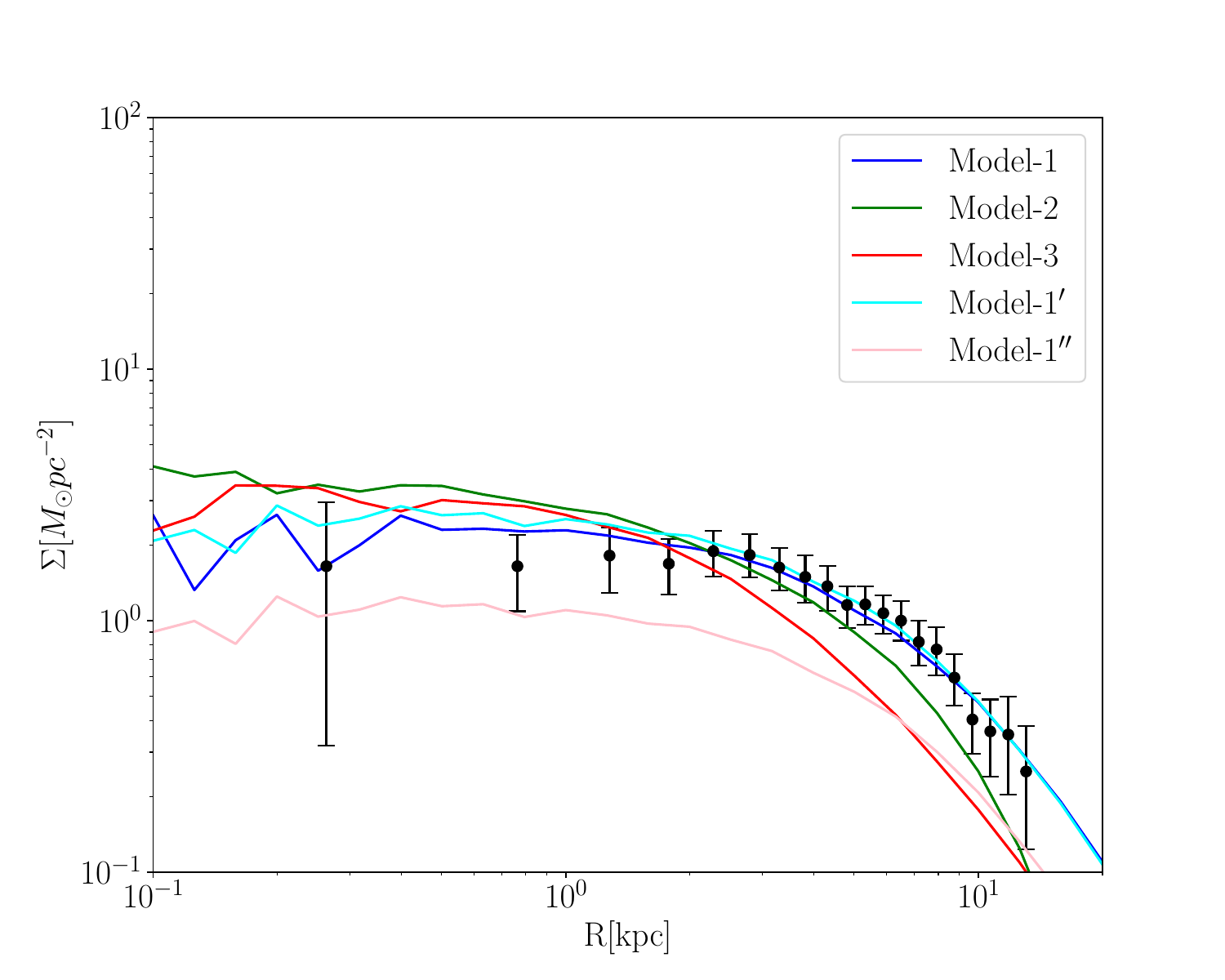}
    \caption{Left panel: initial (dashed lines) and final (solid lines) 2D stellar density profiles of three models under consideration. The color scheme aligns with the colors used in Figure \ref{model}. The black points with error bars represent the observational stellar distribution of Nube \citep{montes2024almost}. The gray shaded region delineates the approximate range covered by the profiles of dwarf galaxies from \cite{Chamba_2020} with stellar masses ($1-5\times 10^8M_\odot$) similar to that of Nube. For comparison, the profile of DF 44, a prototypical UDG, is illustrated by the gray square symbols \citep{montes2024almost}. Right panel: The blue, red, and green solid lines, as well as the black points with error bars, are the same as in the left panel. However, the horizontal axis is set to a logarithmic scale to more intuitively display the profiles at small radii. The cyan and pink solid lines represent the results of Model-$1^\prime$ and -$1^{\prime\prime}$, respectively.}
    \label{result}
\end{figure*}

\subsection{Stellar Distribution in Nube}

To facilitate comparison between the simulation outcomes and observational data, we compute the final 2D stellar density as a function of the distance $R$ from the stellar mass center. The initial and final 2D stellar density profiles of the three models under consideration are depicted in the left panel of Figure \ref{result}, where the lines representing the initial profiles of Model-2 and Model-3 exhibit overlap. The gray shaded region in this panel delineates the approximate range covered by the profiles of dwarf galaxies from \cite{Chamba_2020} with stellar masses ($1-5\times 10^8M_\odot$) similar to that of Nube. Additionally, the gray square symbols represent the profile of DF 44, a prototypical UDG. The comparison between the observational result of Nube (black points with error bars) and the shaded region or the profile of DF 44 intuitively showcases the anomaly of Nube. Our analysis indicates that Model-1 closely matches the observed data, while Model-2 and Model-3 exhibit discrepancies, showing higher and lower densities than observed in the inner and outer regions, respectively. The lower densities in the outer region in Model-2 and Model-3, compared to observational data, cannot be solely explained by the higher densities of these models in the inner region, as the excess stellar mass in the inner region is too small to compensate for the stellar mass deficit in the outer region. Instead, this discrepancy is likely  primarily due to an abundance of stars being pushed into the region with $R\gtrsim 13$ kpc, as evidenced by the bottom row of the left panel in Figure \ref{evolution}. Nevertheless, the surface density in this region is too faint to be detected by current observations. Future observations might reveal this obscured area and enable evaluation of the validity of the FDM hypothesis.

To investigate the impacts of the parameters $m_a$, $k$, and $r_\star$ on the final stellar distribution, the results of two additional models denoted as Model-$1^\prime$ and Model-$1^{\prime\prime}$ are presented in the right panel of Figure \ref{result}. In Model-$1^\prime$,  we maintain the parameters $m_a$, $k$, and $M_\star$ identical to those in Model-1, while replacing $r_\star$ with a value of 1.5 kpc. In Model-$1^{\prime\prime}$, we set $M_\star$ and $r_\star$  equal to the values in Model-2 and Model-3. The result of Model-$1^{\prime\prime}$ is obtained from rescaling the simulation result of Model-$1^\prime$ using the ratio of initial stellar masses. This approach is equivalent to conducting a new simulation separately, as the only difference between these two models lies in the mass assigned to individual stellar particles, while adopting the same number of stellar particles. This difference only influences the normalization of the final stellar density profile, but it does not impact the simulation procedure, where stellar particles are treated as massless. The lower distribution of Model-$1^{\prime\prime}$ compared to the observational data across all ranges can also be attributed to a significant number of stars being pushed beyond 13 kpc.

The comparison between Model-$1$ and Model-$1^{\prime}$ indicates that the variation in $r_\star$ within the range of $1.5-3.0$ kpc has a negligible impact, as previously mentioned. This phenomenon arises from the gradual convergence of stars to a stable density profile under the influence of dynamical heating. For a fixed $M_\star$, any discrepancies induced by differing values of $r_\star$ diminish over time \citep{Yang:2024vgw}. The comparison between Model-$1^{\prime\prime}$ and Model-2 suggests that the dynamical heating effect becomes more pronounced as $m_a$ decreases. This trend aligns with expectations, as smaller $m_a$ enhances the wave effects, at least within the range of $m_a$ under consideration $\sim \mathcal{O}(10^{-23})$ eV. Additionally, the comparison of Model-$1^{\prime\prime}$ and Model-3 reveals that the heating efficiency in the inner region increases with decreasing $k$. This result can be attributed to the reduction in the soliton fraction as $k$ decreases, as shown in Figure \ref{model}. Consequently, the relative ratio of excited states to the soliton rises in the inner region, leading to stronger interference effects and enhanced heating efficiency as $k$ decreases. This understanding is supported by the feature that the amplitude of soliton density oscillation relative to its mean value in Model-$1^{\prime\prime}$ ($92.6\%$) is significantly larger than in Model-3 ($44.6\%$).

\section{Discussions and conclusion\label{Sec4}}

The mechanism utilizing the FDM dynamical heating effect to explain the stellar distribution anomaly in Nube is qualitatively consistent with existing observations of other typical dwarf galaxies or UDGs \citep{Yang:2024vgw}. This consistency arises from the fact that the heating effect is primarily significant in isolated galaxies \citep{Schive_2020} like Nube. Most observed isolated typical dwarf galaxies and UDGs are HI-rich and actively star-forming \citep{Prole_2019}, indicating a much younger age compared to Nube. As a result, there has not been sufficient time for stars in these galaxies to be heated to the extent of becoming as diffuse as those observed in Nube.

Several studies in the literature have suggested a FDM particle mass on the order of $10^{-23}$ eV \citep{Lora_2012,Gonz_lez_Morales_2017,Chiang_2022,Ba_ares_Hern_ndez_2023,ManceraPina:2024ybj}. For instance, this particle mass has been proposed to address various astrophysical phenomena, such as the wide distribution of globular clusters in Fornax \citep{Lora_2012}, the rotation curves of nearby dwarf irregular galaxies \citep{Ba_ares_Hern_ndez_2023}, and extreme galaxies like AGC 114905 \citep{ManceraPina:2024ybj}. 
However, stringent constraints from studies involving the Ly$\alpha$ forest \citep{Rogers_2021}, subhalo mass function \citep{Nadler_2019}, and dynamical heating effect in dwarf galaxies \citep{Marsh_2019,2022PhRvD.106f3517D} suggest a significantly higher FDM particle mass than $\mathcal{O}(10^{-23})$ eV. Nevertheless, it is worth noting that some of these constraints are currently under intense debate regarding many factors, such as uncertainties arising from astrophysical assumptions and data interpretation in the Ly$\alpha$ constraints \citep{Chiang_2022}, or the neglect of tidal suppression effects on the dynamical heating in Segue 1 and 2 \citep{2022PhRvD.106f3517D,Dutta_Chowdhury_2023}. More details on these debates can be found in \cite{Chiang_2022}, \cite{Ba_ares_Hern_ndez_2023}, and \cite{Yang:2024vgw}, emphasizing the need for further research to address these uncertainties. Additionally, the axion mass spectrum expected in string theory covers a wide range \citep{Svrcek:2006yi}, suggesting the possibility of FDM composed of multiple types of particles with different masses. This scenario may relax the current constraints on the FDM particle mass.

One limitation in our simulations is the omission of baryonic feedback effects \citep{Ogiya_2014}, which may be important in the early stages when the stellar distribution in Nube was more concentrated. Incorporating this effect may necessitate a heavier FDM particle. However, relying solely on this effect to explain the characteristics of Nube appears to be quite challenging \citep{montes2024almost}. Another potential effect accounting for the Nube's characteristics is the formation of a cored halo, possibly arising from self-interacting DM \citep{tulin2018dark}. Compared to the cuspy halo in the standard CDM framework, a cored halo features a shallower central gravitational potential, resulting in weaker binding of stars and a more diffuse stellar distribution.  Nevertheless, explaining Nube may require a substantial core size, which in turn would demand a significant self-interaction cross-section among DM particles.

In summary, we employ the dynamical heating effect of FDM to elucidate the extremely diffuse stellar distribution in Nube through simulation. Our findings suggest that an FDM particle mass on the order of $10^{-23}$ eV offers a plausible explanation for the anomaly. A natural consequence of our explanation is the presence of numerous stars in the region $R\gtrsim 13$ kpc. Future observations have the potential to uncover this obscured region and evaluate the validity of the FDM hypothesis. 
\section*{acknowledgments}
This work is supported by the National Natural Science Foundation of China under grant No. 12447105 and 12175248.

\appendix

\section{FDM halo construction\label{App_A}}

We use the eigenstate decomposition method \citep{Yavetz_2022} to construct the initial wave function of FDM in the halo. The time-independent Schr$\ddot{\text{o}}$dinger equation under the potential $\Phi_\text{in}(r)$, which is determined by the target profile $\rho_\text{in}(r)$, is expressed as 
\begin{equation}
    -\frac{\hbar^2}{2m_a}\bm{\nabla}^2\Psi_{nlm}(\mathbf{x})+m_a\Phi_\text{in}(r)\Psi_{nlm}(\mathbf{x})=E_{nl}\Psi_{nlm}(\mathbf{x}).
    \label{eigen}
\end{equation}
Substituting $\Psi_{nlm}(\mathbf{x})=R_{nl}(r)Y^m_l(\theta,\varphi)$ into Equation (\ref{eigen}) and defining an auxiliary function $u_{nl}\equiv rR_{nl}$, we derive the equation governing the initial radial wave function as 
\begin{equation}
\begin{aligned}
    &-\frac{\hbar^2}{2m_a}\frac{d^2u_{nl}}{dr^2}+\left[\frac{\hbar^2}{2m_a}\frac{l(l+1)}{r^2}+m_a\Phi_\text{in}\right]u_{nl}=E_{nl}u_{nl},
\end{aligned}
\label{u_nl}
\end{equation}
where $E_{nl}$ is the eigenvalue associated with the eigenstate.
The normalization condition of $u_{nl}(r)$ and boundary conditions of Equation (\ref{u_nl}) are specified as
\begin{equation}
    \int_0^\infty u^2_{nl}(r)dr=1,\,u_{nl}(0)=0,\,\lim_{r\to\infty}u(r)=0.
    \label{bound_u}
\end{equation}

To simplify the solution procedure, we nondimensionalize Equation (\ref{u_nl}) using the same length, time, and mass scales as those used in the FDM wave function evolution \citep{Edwards_2018}. These scales are $\mathcal{L}\simeq 121\left(10^{-23}\text{eV}/m_a\right)^{1/2}\text{kpc},\mathcal{T}\simeq 75.5\text{Gyr}$, and $\mathcal{M}\simeq 7\times 10^7\left(10^{-23}\text{eV}/m_a\right)^{3/2}M_\odot$.
Introducing another auxiliary function $\tilde{v}_{nl}$ enables us to reformulate Equation (\ref{u_nl}) as a system of two first-order differential equations,
\begin{equation}
     \frac{d\tilde{u}_{nl}}{d\tilde{r}}=\tilde{v}_{nl},\,\frac{d\tilde{v}_{nl}}{d\tilde{r}}=2\left[\frac{l(l+1)}{2\tilde{r}^2}+\widetilde{\Phi}_\text{in}-\widetilde{E}_{nl}\right]\tilde{u}_{nl},
    \label{u_v}
\end{equation}
where $\tilde{u}_{nl}\equiv \mathcal{T}\sqrt{m_aG}u_{nl}/\mathcal{L}u_{nl},\tilde{r}\equiv r/\mathcal{L},\widetilde{\Phi}_\text{in}=m_a\mathcal{T}\Phi_\text{in}/\hbar$, and $\widetilde{E}_{nl}\equiv \mathcal{T}E_{nl}/\hbar$. The normalization condition and boundary conditions of $u_{nl}$ can be equivalently expressed in terms of those for   $\tilde{u}_{nl}$ as
\begin{equation}
\int_0^\infty\tilde{u}^2_{nl}d\tilde{r}=\frac{\mathcal{L}^3}{m_aG\mathcal{T}^2},\,\tilde{u}_{nl}(0)=0,\, \lim_{\tilde{r}\to\infty}\tilde{u}_{nl}(\tilde{r})=0.
    \label{b_u}
\end{equation}
Equations (\ref{u_v}) and (\ref{b_u}) can be viewed as an eigenvalue problem. For a given value of $\widetilde{E}_{nl}$, the boundary condition at $\tilde{r}=\infty$ and the normalization condition together uniquely determine a solution to Equation (\ref{u_v}). However, the boundary condition at $\tilde{r}=0$ is only satisfied for certain specific values of $\widetilde{E}_{nl}$, which correspond to the eigenvalues of Equation (\ref{u_v}).  
 
We adopt the shooting method to solve this eigenvalue problem. Specifically, for a given set of $l$ and $\widetilde{E}_{nl}$ values, we numerically solve Equation (\ref{u_v}) within a finite grid spanning from $\tilde{r}_1=r_1/\mathcal{L}$ to $\tilde{r}_2=r_2/\mathcal{L}$, where $r_1$ and $r_2$ are selected as $0.01$ kpc and $4\,r_\text{vir}$, respectively. To determine the correct $\widetilde{E}_{nl}$ as an eigenvalue, we utilize the bisection method to iteratively adjust $\widetilde{E}_{nl}$ such that the boundary conditions in Equation (\ref{b_u}) are satisfied. Once the eigenvalue $\widetilde{E}_{nl}$ is determined, we normalize the corresponding numerical form of $\tilde{u}_{nl}$ based on the normalization condition in Equation (\ref{b_u}). Through some trivial transformations, we can obtain the corresponding $R_{nl}(r)$ for this eigenstate. We have verified that the solutions remain almost unchanged even when employing a broader region of $\left[\tilde{r}_1,\tilde{r}_2\right]$ for numerical computations.

In practical operations, we restrict our consideration to the eigenstates $\Psi_{nlm}(\mathbf{x})$ with eigenenergies below a maximum energy cutoff $E_c$, which  is set to the energy of a particle in a circular orbit at the virial radius. After obtaining $\Psi_{nlm}(\mathbf{x})$, the initial wave function $\psi(0,\mathbf{x})$ for our simulations can be written as a linear combination of these eigenstates:
\begin{equation}
    \psi(0,\mathbf{x})=\sum_{nlm}a_{nlm}\Psi_{nlm}(\mathbf{x}).
    \label{psi_1}
\end{equation}
The total number of eigenstates obtained for our three models in Table \ref{Tab1} of the main text are 13469, 228607, and 7719, respectively. To simplify the subsequent analysis, we omit the $m$ dependence of the coefficients' amplitude $|a_{nlm}|$,  representing them as $|a_{nl}|$, while retaining the $m$ dependence of their phases. This reduces the number of free parameters to 533, 3489, and 373 for our three models, respectively. 

Then, the next step involves adjusting the magnitudes $|a_{nl}|$ to ensure that the random phase averaged output profile $\rho_\text{out}(r)=\frac{m_a}{4\pi}\sum_{nl}(2l+1)|a_{nl}|^2R^2_{nl}(r)$ matches the input target profile $\rho_\text{in}(r)$. We further reduce the number of free parameters $|a_{nl}|$ by dividing the energy range from the minimum value of the gravitational potential energy to the selected maximum energy cutoff $E_c$ into 60 bins uniformly and assuming that the coefficients $|a_{nl}|$ of the eigenstates within the same bin of the eigenenergy are equal. Since the form of $\rho_\text{out}(r)$ is a linear combination of functions $R^2_{nl}(r)$, where the coefficients of the combination are proportional to $|a_{nl}|^2$, we can utilize the nonnegative least squares method to determine the optimal values of $|a_{nl}|^2$, which improves the fitting speed greatly. The maximum radius adopted for fitting the input target profile is set to $1.2~r_\text{vir}$. Finally, we assign a random phase, which is dependent on $m$, to each $a_{nlm}$ to obtain the initial wave function $\psi(0,\mathbf{x})$. 

\section{Evolution of the system\label{App_B}}

We utilize the \textsc{PyUltraLight} package \citep{Edwards_2018}, which adopts the pseudospectral method, to evolve the FDM wave function satisfying the SP equations. By using the length, time, and mass scales defined in Appendix \ref{App_A}, we can nondimensionalize the time-dependent Schr$\ddot{\text{o}}$dinger equation as follows
\begin{equation}
    i\frac{\partial}{\partial\tilde{t}}\widetilde{\psi}(\tilde{t},\mathbf{\tilde{x}})=-\frac{1}{2}\widetilde{\bm{\nabla}}^{2}\widetilde{\psi}(\tilde{t},\mathbf{\tilde{x}})+\widetilde{\Phi}(\tilde{t},\mathbf{\tilde{x}})\widetilde{\psi}(\tilde{t},\mathbf{\tilde{x}}),
\end{equation}
where $\widetilde{\psi}\equiv \mathcal{T}\sqrt{m_aG}\psi,\widetilde{\Phi}\equiv m_a\mathcal{T}\Phi/\hbar,\tilde{t}=t/\mathcal{T}$, and $\mathbf{\tilde{x}}\equiv\mathbf{x}/\mathcal{L}$. The dimensionless wave function is evolved using the unitary time evolution operator, with certain operations being more conveniently performed in Fourier space. The evolution is given by
\begin{equation}
\begin{aligned}
    &\widetilde{\psi}(\tilde{t}+\Delta\tilde{t}_\text{FDM},\tilde{\mathbf{x}})=\exp\left[-\frac{i\Delta\tilde{t}_\text{FDM}}{2}\widetilde{\Phi}(\tilde{t}+\Delta\tilde{t}_\text{FDM},\tilde{\mathbf{x}})\right]\\
    &\hspace{2.3cm}\times\mathcal{F}^{-1}\left\{\exp\left(-\frac{i\Delta\tilde{t}_\text{FDM}}{2}k^2\right)\right.\\
    &\hspace{2.5cm}\left.\mathcal{F}\left[\exp\left[-\frac{i\Delta\tilde{t}_\text{FDM}}{2}\widetilde{\Phi}(\tilde{t},\tilde{\mathbf{x}})\right]\widetilde{\psi}(\tilde{t},\tilde{\mathbf{x}})\right]\right\},
\end{aligned}
\end{equation}
where $\mathcal{F}$ and $\mathcal{F}^{-1}$ represent the Fourier and inverse Fourier transformations, respectively. The dimensionless gravitational field generated by FDM at dimensionless time $\tilde{t}$ can be calculated as
\begin{equation}
    \widetilde{\Phi}(\tilde{t},\tilde{\mathbf{x}})=\mathcal{F}^{-1}\left\{-\frac{1}{k^2}\mathcal{F}\left[4\pi\left|\widetilde{\psi}(\tilde{t},\tilde{\mathbf{x}})\right|^2\right]\right\}.
\end{equation}
With this methodology, we can obtain the dimensional wave function $\psi(t,\mathbf{x})$ and gravitational field $\Phi(t,\mathbf{x})$ at grid points at time $t$. 

To determine the acceleration experienced by a star particle at any position within the simulation box, we interpolate the gravitational field at grid points to obtain a continuous field throughout the box. The acceleration of a star at position $\mathbf{x}(t)$ can  be calculated using Newton's second law $\mathbf{a}(t)=-\bm{\nabla}\Phi(t,\mathbf{x}(t))$. Subsequently, the position and velocity of this star particle at $t+\Delta t_\star$ can be determined based on $\mathbf{x}(t),\mathbf{v}(t)$ and $\mathbf{a}(t)$. To enhance the accuracy of particle evolution, we employ the fourth-order Runge-Kutta method to calculate the updating of each star's position and velocity.

\bibliography{Refs}
\bibliographystyle{aasjournal}

\end{document}